\documentclass[11pt,fleqn]{article}

\usepackage{amsfonts,amssymb,latexsym,cite}
\usepackage{epsf,graphicx}
\usepackage{amsmath, amssymb}
\usepackage{epstopdf}
\usepackage{eurosym}

\usepackage{amsfonts,amssymb,cite}
\usepackage{epsf,graphicx}

\def\noi{\noindent}

\newcommand{\Title}[1]{\noi {{\Large\bf #1}}\\[1ex]}

\def\Aunames#1{\noi{\bf #1}}
\def\auth#1{${}^{#1}$}
\def\Addresses#1{\medskip\noi \protect
	\begin{description}\itemsep -3pt {\it #1} \end{description}}
\def\addr#1#2{\item[${}^{#1}$]{\it #2}}

\newcommand{\Abstract}[1]{\vskip 2mm \begin{center}
        \parbox{16.4cm}{\small\noi #1} \end{center}\medskip}

\def\email#1#2{\footnotetext[#1]{e-mail: #2}\addtocounter{footnote}{1}}

\def\nqq{\hspace*{-2em}}
\def\nhq{\hspace*{-0.5em}}

\def\inch{\hspace*{1in}}

\def\Jl#1#2{#1 {\bf #2},\ }

\def\ApJ#1 {\Jl{Astroph. J.}{#1}}
\def\CQG#1 {\Jl{Class. Quantum Grav.}{#1}}
\def\DAN#1 {\Jl{Dokl. AN SSSR}{#1}}
\def\GC#1 {\Jl{Grav. Cosmol.}{#1}}
\def\GRG#1 {\Jl{Gen. Rel. Grav.}{#1}}
\def\JETF#1 {\Jl{Zh. Eksp. Teor. Fiz.}{#1}}
\def\JETP#1 {\Jl{Sov. Phys. JETP}{#1}}
\def\JHEP#1 {\Jl{JHEP}{#1}}
\def\JMP#1 {\Jl{J. Math. Phys.}{#1}}
\def\NPB#1 {\Jl{Nucl. Phys. B}{#1}}
\def\NP#1 {\Jl{Nucl. Phys.}{#1}}
\def\PLA#1 {\Jl{Phys. Lett. A}{#1}}
\def\PLB#1 {\Jl{Phys. Lett. B}{#1}}
\def\PRD#1 {\Jl{Phys. Rev. D}{#1}}
\def\PRL#1 {\Jl{Phys. Rev. Lett.}{#1}}

\def\al{&\nhq}
\def\lal{&&\nqq {}}
\def\eq{Eq.\,}
\def\eqs{Eqs.\,}
\def\beq{\begin{equation}}
\def\eeq{\end{equation}}
\def\bear{\begin{eqnarray}}
\def\bearr{\begin{eqnarray} \lal}
\def\ear{\end{eqnarray}}
\def\earn{\nonumber \end{eqnarray}}

\def\nnn{\nonumber\\ \lal }

\def\yy{\\[5pt] {}}
\def\yyy{\\[5pt] \lal }
\def\eql{\al =\al}

\def\dst{\displaystyle}
\def\tst{\textstyle}
\def\fracd#1#2{{\dst\frac{#1}{#2}}}
\def\fract#1#2{{\tst\frac{#1}{#2}}}
\def\Half{{\fracd{1}{2}}}
\def\half{{\fract{1}{2}}}

\def\d{\partial}

\def\const{{\rm const}}

\def\mN{_\mu^\nu}
\def\nM{_\nu^\mu}

\def\ssph{static, spherically symmetric}
\def\asflat{asymptotically flat}

\def\R{{\mathbb R}}

\begin{document}
\twocolumn[
\vspace{15mm}

\Title{On horizons and wormholes in k-essence theories}

\Aunames{K. A. Bronnikov,\auth{a,b,c,1}
	 J. C. Fabris,\auth{c,d,2}
	 and D. C. Rodrigues\auth{d}
	 }

\Addresses{
\addr a {VNIIMS, Ozyornaya ul. 46, Moscow 119361, Russia}
\addr b {Institute of Gravitation and Cosmology, PFUR,
         ul. Miklukho-Maklaya 6, Moscow 117198, Russia}
\addr c {National Research Nuclear University ``MEPhI'',
	 Kashirskoe sh. 31, Moscow 115409, Russia}
\addr d {Universidade Federal do Esp\'{\i}rito Santo,
 	 Vit\'oria, ES, CEP29075-910, Brazil}
	}


\Abstract
  {We study the properties of possible static, spherically symmetric
   configurations in k-essence theories with the Lagrangian
   functions of the form $F(X)$, $X \equiv \phi_{,\alpha} \phi^{,\alpha}$.
   A no-go theorem has been proved, claiming that a possible black-hole-like
   Killing horizon of finite radius cannot exist if the function $F(X)$ is
   required to have a finite derivative $dF/dX$. Two exact solutions are
   obtained for special cases of k-essence: one for $F(X) =F_0 X^{1/3}$,
   another for $F(X) = F_0 |X|^{1/2} - 2 \Lambda$, where $F_0$ and $\Lambda$
   are constants. Both solutions contain horizons, are not asymptotically
   flat, and provide illustrations for the obtained no-go theorem.
   The first solution may be interpreted as describing a black hole in an
   asymptotically singular space-time, while in the second solution two
   horizons of infinite area are connected by a wormhole.
   }

] 
\email 1 {kb20@yandex.ru}
\email 2 {fabris@pq.cnpq.br}

\section{Introduction}

  A great number of modifications and extensions of the century-old general
  relativity (GR) theory have been proposed since its formulation in 1915,
  and, which is surprising and remarkable, they are continuing to emerge
  nowadays. Such new proposals are motivated by many reasons, among them are
  the old problem of unifying gravity with other physical interactions and
  the difficulties in attempts to quantize GR. One should also mention
  two main problems concerning classical GR itself: the existence of
  singularities in the most physically relevant solutions of GR, and the
  necessity to introduce unknown forms of matter in order to explain the
  main features of the observed universe. The modifications evoked in the
  literature can be divided into two large classes. In the first one, the
  geometric sector is generalized: it includes, in particular, $f(R)$
  thories, multidimensional theories and non-Riemannian geometries. The
  second class involves new fundamental, non-geometric fields with
  nonstandard structure. To the second class one may attribute scalar-tensor
  theories, Galileons and Horndesky theories and others. It is frequently
  possible to establish a connection between the two approaches.

  The k-essence theories \cite{mukha1} evidently belong to those with
  nonstandard fundamental fields coupled to gravity. This class is based on
  a possible non-standard form of the kinetic term of a scalar field.
  It was for the first time suggested in \cite{mukha2} in order to have an
  inflationary model driven by the kinetic term instead of the potential.
  But soon after this idea was applied to explain the present stage
  of accelerated expansion of the Universe \cite{mukha3}. It is interesting
  to note that a k-essence structure also appears in string theories
  as, for example, in the Dirac-Born-Infeld action, where the kinetic
  term of the scalar field has a structure similar to the Maxwell-like term
  in Born-Infeld electrodynamics \cite{dbi}.

  The k-essence theories can be defined by the following most general
  Lagrangian:
\bearr 			\label{L1}
	{\cal L} = \sqrt{-g}[R - F(X,\phi)],
\ear
  with
\bear
	X = \eta\phi_{;\rho}\phi^{;\rho},
\ear
  where $\eta = \pm 1$ can be used to make $X$ positive in the cases
  like general power-law dependence, ill-defined for negative quantities.
  There are other, more special presentations of k-essence Lagrangians,
  for example,
\bear \label{L2}
	{\cal L} = \sqrt{-g}[R - f(X) + 2V(\phi)],
\ear
  separating the kinetic and potential terms.

  While many studies have been carried out in the context of cosmology for
  k-essence theories, a much smaller effort was applied to study the possible
  effect of k-essence on the structure and properties of local objects,
  like, for instance, black holes and wormholes. The aim of this paper is
  to consider possible static, spherically symmetric configurations in
  theories defined by the Lagrangian (\ref{L1}).

  Although the field equations are written in full generality for the
  Lagrangian (\ref{L1}) (Section 2), the results obtained here actually
  concern the more specific form (\ref{L2}). The complexity of the equations
  prevents us to find more or less general explicit solutions. However, it
  has been possible to prove a general no-go theorem which states that, in
  the absence of a $\phi$-dependent potential term in (\ref{L2}), only
  horizons inherent to cold black holes \cite{k1,k2} can appear in these
  theories, similarly to what happens in scalar-tensor theories (Section 3).
  A cold black hole is a term coined in \cite{k1,k2} to designate
  asymptotically flat static spherical solutions where the horizon surface
  is infinite. The surface gravity of such black holes is zero, implying a
  zero Hawking temperature.  However, the tidal forces acting on extended
  test bodies are infinite at the horizon. It turns out that in
  scalar-tensor theories like the Brans-Dicke theory, in the absence of a
  potential term in the Lagrangian, the scalar-vacuum solutions in general
  contain naked singularities and, in some special cases, cold black hole
  solutions are possible. We will show that, again in the absence of a
  potential, only cold black hole horizons are possible in k-essence
  theories.

  Further on we obtain two special exact solutions, one for $V=0$ and
  $f(X)\sim X^{1/3}$ (Section 4), another for $f(X) \sim |X|^{1/2}$ in the
  presence of a cosmological constant (Section 5), and briefly describe
  their properties. Section 6 contains some final remarks.

\section{Basic equations}

  Variation of the Lagrangian (\ref{L1}) with respect to the metric and
  the scalar field leads to the field equations
\bearr                                   		    \label{EE}
	G_\mu^\nu = -T\mN [\phi],
\yyy
	T\mN [\phi] \equiv \eta F_X \phi_\mu \phi^\nu
		- \frac{1}{2} \delta\mN F,
\yyy   					      		 \label{eq-phi}
       \eta\nabla_\alpha (F_X \phi^\alpha) - \Half\, F_\phi = 0,
\ear
  where $G\mN$ is the Einstein tensor, $F_X = \d F/\d X$, $F_\phi = \d
  F/\d\phi$, and $\phi_\mu = \d_\mu \phi$.

  Now, consider a general \ssph\ metric
\beq                                              \label{ds}
	ds^2 = e^{2\gamma(u)}dt^2 - e^{2\alpha(u)}du^2
			- e^{2\beta(u)} d\Omega^2
\eeq
  ($d\Omega^2 = d\theta^2 + \sin^2 \theta d\varphi^2$ is the metric on
  a unit sphere) with an arbitrary radial coordinate $u$, and $\phi =
  \phi(u)$. Then, in the general case (\ref{L1}), the stress-energy tensor
  (SET) $T\mN$ has the following nonzero components:
\bearr                                                       \label{SET}
	T^0_0 = T^2_2 = T^3_3 = - F/2,
\nnn
	T^1_1 = -F/2 - \eta F_X e^{-2\alpha} \phi'^2,
\ear
  where the prime denotes $d/du$. In the case under consideration, $X = -\eta
  e^{-2\alpha}\phi'^2$, and to make $X$ positive, in what follows we put
  $\eta = -1$ (unless otherwise indicated).

  The scalar field equation and the nontrivial components of the Einstein
  equations can be written as follows:
\bearr                                                        \label{eq-s}
	2\Big(F_X e^{-\alpha + 2\beta + \gamma} \phi'\Big)'
	      - e^{\alpha + 2\beta + \gamma} F_\phi = 0,
\yyy                                                            \label{00}
	\gamma'' + \gamma' (2\beta'+\gamma'-\alpha')
					= \half e^{2\alpha} (F - X F_X),
\yyy
	-e^{2\alpha - 2\beta} +                                 \label{22}
	\beta'' + \beta' (2\beta'+\gamma'-\alpha')
\nnn \inch
	     	= \half e^{2\alpha} (F - X F_X),
\yyy
	-e^{-2\beta} + e^{-2\alpha} \beta'(\beta'+2\gamma')   \label{11}
					= \half\,F - X F_X,
\ear
  where (\ref{11}) (the component $G^1_1 = \ldots$) is a first integral of
  the other equations.

  In particular, we will use the so-called quasiglobal radial coordinate
  \cite{BR-book} $u =: x$ specified by the condition $\alpha(u) + \gamma(u)
  =0$, it is especially convenient for considering Killing horizons
  which are then described as regular zeros of the function
  $A(x) = e^{2\gamma} = e^{-2\alpha}$. The metric has the form
\beq                                                         \label{ds-q}
	ds^2 = A(x) dt^2 - \frac{dx^2}{A(x)} - r^2(x) d\Omega^2.
\eeq
  In these coordinates, two combinations
  of \eqs (\ref{00})--(\ref{11}) take rather a simple form:
\bearr
	2 A\,\frac{r''}{r} = X\,F_X,                  	\label{01}
\yyy
	A'' r^2 - A (r^2)'' = -2,                       \label{02}
\ear
  where now the primes stand for $d/dx$. The other two equations,
  (\ref{eq-s}) and (\ref{11}), are rewritten as
\bearr                                                    \label{s-q}
	2 (F_X\,A r^2 \phi')' - r^2 F_\phi =0,
\yyy                                                      \label{11-q}
	\frac{1}{r^2}(-1 + A' r r' + Ar'^2) = \Half F - X F_X.
\ear
  Equations (\ref{01}) and (\ref{02}) are the combinations
  $G^0_0 -G^1_1 = \ldots$ and $G^0_0 - G^2_2 = 0$.
  Equation (\ref{02}) can be once easily integrated:
\bear                                      		\label{B'}
	  B'(x) \equiv \biggl(\frac{A}{r^2}\biggr)' = \frac{6m - 2x}{r^4},
\ear
  where the constant $m$ has the meaning of the Schwarzschild mass if
  the metric is \asflat\ as $x\to\infty$.

  An important point connected with \eq (\ref{01}) is that it relates the
  sign of the difference $T^0_0 - T^1_1 = - X F_X = \rho + p$ (in the
  standard perfect-fluid notation) with the quantity $r''$ \cite{BR-book}.
  Namely, if $r'' < 0$, then $T^0_0 - T^1_1 > 0$, and the Null Energy
  Condition (NEC) is fulfilled; on the contrary, if $r'' > 0$, this
  condition is violated, and, in particular, wormhole throats are possible.
  It is clear that the general Lagrangian (\ref{L1}) or even (\ref{L2}) make
  possible any sign of $X F_X$ and hence $r''$.

\section{A no-go theorem}

  The structure of the SET (\ref{SET}) leads to an important statement about
  possible horizons in space-times with the metric (\ref{ds}) even in the
  general case (\ref{L1}) ({\it the Global Structure Theorem\/})
  \cite{kb-01}: there can be at most two simple (Schwarzschild-like)
  horizons at finite radius $r = e^\beta$, and no more than one such horizon
  if the space-time is \asflat. This result directly follows from the
  equality $T^0_0 = T^2_2$, leading to \eq (\ref{02}) that does not contain
  any functions of $\phi$.

  Indeed, as already mentioned, horizons are described by regular zeros
  of the function $A(x)$ or, equivalently, $B(x) = A/r^2$ (provided that $r$
  is finite). Meanwhile, it follows from (\ref{02}) that the function $B(x)$
  cannot have a regular minimum, therefore, once having become negative,
  $B(x)$ never returns to zero.

  There are many other results of interest concerning the possible existence
  or non-existence of horizons in configurations with scalar fields. Let us
  mention, for instance, the no-hair theorem by Adler and Pearson
  \cite{ad-pearson}, claiming that there cannot be \asflat\ black holes with
  a nontrivial scalar field in the case $F = X - 2V(\phi)$ with $\eta
  =1$ (a normal, non-phantom scalar field) and nonnegative potentials $V$.
  It was generalized in \cite{kb-02} to certain multiscalar and
  multidimensional space-times.

  Here we will obtain one more ``no-hair'' result concerning the important
  family of Lagrangians (\ref{L1}), those with $F= f(X)$.

  With $F = f(X)$, \eq (\ref{eq-s}) is integrated giving
\bear                                                     \label{phi-C}
          f_X e^{-\alpha+2\beta+\gamma} \phi' = C = \const,
\ear
  or, if we again use the quasiglobal coordinate $u=x$,
\beq
	  A f_X \phi' = C/r^2.                            \label{phi-q}
\eeq
  and \eq (\ref{01}) can be rewritten as
\beq                                           \label{01a}
       2 \frac{r''}{r} = - \phi'{}^2 f_X = - \frac{C^2}{A^2 r^4 f_X},
\eeq

  Let us look whether or not the system admits a Killing horizon like
  event horizons of static black holes. From the above-mentioned Global
  Structure Theorem \cite{kb-01} it follows that if $A(x) > 0$ in
  some range of $x$ (where the metric is really static), such a horizon
  can only be first-order, such that the function $A(x)$ behaves as $A \sim
  x-x_h$.

  Looking at (\ref{phi-q}), we see that $A\to 0$ at finite $r$ implies
  $f_X \phi' \to \infty$ (excluding the trivial case $C=0$). On the other
  hand, again due to (\ref{phi-q}), the expression $X f_X$, which enters
  into the SET, is equal to $C\phi'/r^2$. Therefore finiteness of the SET
  components $T\mN$, necessary for space-time regularity\footnote
  	{The expression $T\mN T\nM$ is a sum of squares and is proportional
	to the Ricci tensor invariant $R\mN R\nM$. Therefore, for finiteness
	of this curvature invariant it is necessary that each term in this
	sum of squares be finite.}
  implies $|\phi'| < \infty$, hence a horizon requires $f_X \to \infty$.

  Thus we have obtained the following no-go theorem:
  {\it The existence of a black-hole-like Killing horizon at finite $r$
  is incompatible with a regular function $f(X)$.}

  Meanwhile, what we have called a cold black hole \cite{k1,k2}, with a
  horizon of infinite area ($r = \infty$), still remains possible if $Ar^2$
  tends to a finite constant.

\section{Special solution 1}

  It happens that \eqs (\ref{eq-s})--(\ref{11}), or equivalently
  (\ref{01})--(\ref{11-q}), are quite hard to solve, even in the
  comparatively simple case
\beq                                                        \label{F-n}
	F(X) = F_0 X^n, \qquad F_0,\ n = \const.
\eeq
  To our knowledge, the only thus far existing example of an exact solution
  is the well-known case of a linear massless scalar field, $n=1$, with
  Fisher's solution \cite{fisher} and its phantom (``anti-Fisher'')
  counterpart leading to the simplest wormhole solutions
  \cite{h_ellis, kb-73}. We will give here one more example,
  with $n=1/3$, which, although looks somewhat exotic, still illustrates
  the no-go theorem obtained in the previous section and has some features of
  interest.

  Let us use, as before, the quasiglobal coordinate $u=x$, corresponding to
  the metric (\ref{ds-q}). Under the assumption (\ref{F-n}), \eq
  (\ref{phi-q}) leads to the following expression for $\phi'$:
\beq                                                           \label{phi'}
	\phi' = \left[ \frac{C}{n F_0 A^n r^2} \right]^{1/(2n-1)}.
\eeq
  Substituting it into (\ref{01}), we obtain
\beq
	2 \frac{r''}{r} =
	\biggl[ (nF_0)^{-1} A^{1-3n} C^{2n} r^{-4n}\biggr]^{1/(2n-1)}.
\eeq
  We see that the function $A(x)$ drops out from this equation in the case
  $n = 1/3$, which then leads to the equation
\beq
	r'' = 3K r^5, \qquad 3K := F_0^3 /(54 C^2).
\eeq
  whose first integral is
\beq
	r'^2 = K r^6 + K_1, \qquad K_1 = \const.
\eeq
  This equation can be further integrated, but with $K_1 \ne 0$ one obtains
  very bulky expressions with elliptic integrals which will not be
  considered. Assuming $K_1 =0$, we easily obtain without loss of generality
\beq                                                        \label{r1}
	r(x) = \frac{1}{k \sqrt{x}}, \quad\
	k =\sqrt{2\sqrt{K}} = \frac{2^{1/4}}{3}\frac{F_0^{3/4}}{\sqrt{|C|}},
\eeq
  where we have suppressed the emerging integration constant by choosing the
  zero point of $x$. Thus the solution is defined at $x > 0$.

  From (\ref{r1}) it follows that $r'' = 3/(4k x^{5/2}) > 0$, it means that
  the NEC is violated, and our k-essence field is of phantom nature.

  It is straightforward to find the expressions for
  $B(x)$ and $\phi'(x)$ from (\ref{B'}) and (\ref{phi'}):
\bear                                                          \label{B}
	B(x) \eql B_0 + k^4 \Big(2m x^3 - \Half x^4\Big),
\yy                                                            \label{phi1}
	\phi'(x) \eql \biggl(\frac{F_0}{3C}\biggr)^3 B r^8.
\ear

  A further substitution of these quantities to \eq (\ref{11-q})
  should verify the correctness of the solution and maybe lead to a relation
  between its integration constants. Doing so, we find in the left-hand side
  of (\ref{11-q})
\beq
	G^1_1 = -3 k^2 m + \frac{3B}{4k^2 x^3},
\eeq
  whereas the right-hand side, $-T^1_1$, contains only the second term of
  this expression. We conclude that this solution only exists with $m =0$.
  The resulting metric has the form
\bearr                               			\label{ds-1}
       ds^2 = \Big(\frac{B_0}{k^2 x} - \frac{k^2}{2}x^3\Big) dt^2
	-\Big(\frac{B_0}{k^2 x} - \frac{k^2}{2}x^3\Big)^{-1} dx^2
\nnn \inch
	- \frac{1}{k^2 x} d\Omega^2.
\ear
  If $B_0 \leq 0$, the function $A(x)$ is negative, and the metric
  describes a particular Kantowski-Sachs cosmological model. If $B_0 > 0$,
  the metric is static at $x < x_h = (2B_0)^{1/4}/k$, has a horizon at
  $x = x_h$ and is cosmological at larger $x$. Although the metric is
  perfectly regular in the whole range $x \in \R_+$, the original function
  $F(X)$ in the Lagrangian is singular at the horizon. Indeed, we easily
  verify that $F_X = (F_0/3) X^{-2/3} \propto 1/A^2$, it is infinite at a
  horizon where $A=0$ and $X=0$, in full agreement with the above no-go
  theorem. But of interest is the very existence of a regular metric in the
  presence of a singular function $F(X)$ in the Lagrangian.

  Another observation of interest is the negative sign of $X = A\phi'^2$ in
  the T-region $x > x_h$. The existence and regular behavior of the solution
  in this region is evidently related to the odd denominator in the exponent
  1/3, owing to which we simply have there $F(X) < 0$, whereas for general
  $n$ the power-law function $X^n$ is ill-defined.

  With any $B_0\ne 0$, the metric (\ref{ds-1}) has singularities
  both at $x=0$ where $r = \infty$ and at $r = \infty$ where $r =0$: at
  both ends, the scalar field $\phi$ and the Kretschmnn scalar
  $R_{\mu\nu\rho\sigma} R^{\mu\nu\rho\sigma}$ tend to infinity. The
  Carter-Penrose diagram in the case $B_0 > 0$ (Fig.\,1) looks like that for
  de Sitter space-time. However, unlike the latter, the nonstatic T-region
  here corresponds to smaller radii $r(x)$ than the static region, just as
  in black hole space-times; in addition, now all sides of the square in the
  diagram correspond to singulariries. One can conclude that the solution
  describes a black hole in space-time with a singular asymptotic.

\begin{figure}[h]
\centering
\includegraphics[width=5cm]{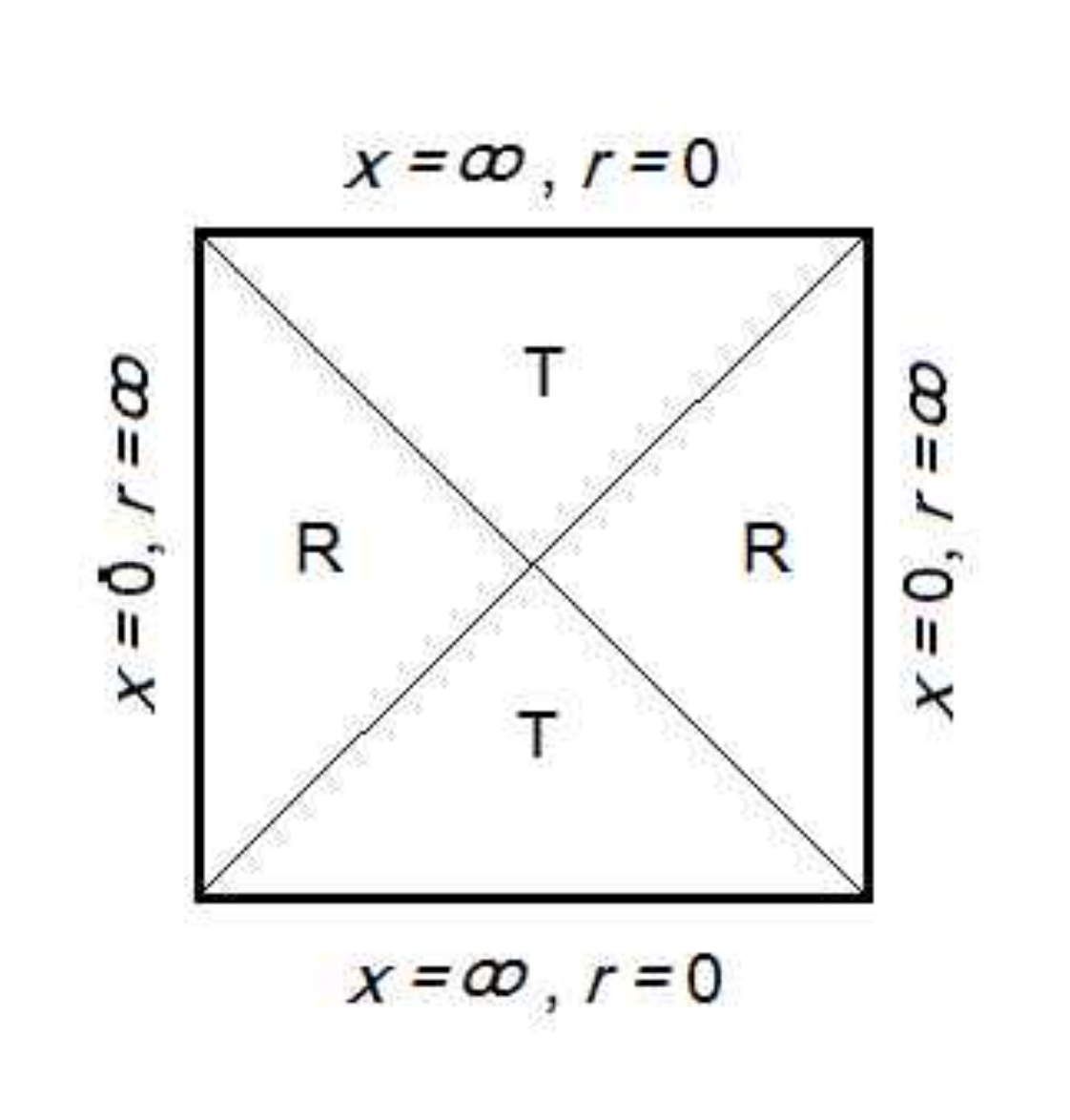}
\caption{\small
	Carter-Penrose diagram for solution 1 with $B_0 > 0$. The
	letters R and T correspond to static and nonstatic regions,
	respectively.} \label{f1}
\end{figure}

\section{Special solution 2}

  Let us now consider the special case of the Lagrangian (\ref{L2}) with
\beq
	F = F(X) = F_0 X^n -2 \Lambda,\quad\ F_0, \Lambda = \const.
\eeq
  As before, we put $\eta =-1$, so that, in terms of the general form
  of the metric (\ref{ds}), $X = e^{-2\alpha} \phi'{}^2 > 0$.
  Note that this specific choice of $F(X)$ has found some interesting
  applications in cosmology \cite{sahni}. Here we consider it for \ssph\
  configurations.

  This time we will use the harmonic coordinate condition \cite{kb-73}
\beq                                                   \label{harm}
  	\alpha = \gamma + 2\beta,
\eeq
  Then \eq (\ref{eq-s}) leads to
\beq
	\Big[n e^{2(1-n)\alpha} \phi'{}^{2n-1}\Big]' =0
\eeq
  The derivative $\phi'$ drops out from this equation in the case $n = 1/2$,
  and it then follows $\alpha = \const$. It makes sense to put $e^\alpha
  = a^2$ where $a$ is a constant with the dimension of length specifying a
  length scale (note that due to the coordinate condition (\ref{harm}) the
  coordinate $u$ has the dimension of 1/length). With (\ref{harm}) we thus
  obtain $e^\gamma = a^2 e^{-2\beta}$, and a difference of \eqs (\ref{00})
  and (\ref{22}) takes the easily integrable Liouville form $3\beta'' = a^4
  e^{-2\beta}$ whose integration gives
\beq                                                    \label{beta}
	r(u) \equiv e^{\beta} = \frac{a^2}{\sqrt{3}b}\cosh (bu)
		  = \frac{b}{\sqrt{3}} \cosh (bu),
\eeq
  where $b > 0$ is an integration constant with the dimension of length
  (one more integration constant is suppressed by choosing the zero point of
  $u$); furthermore, without loss of generality, we have put the length
  scale $a$ equal to $b$. Substituting (\ref{beta}) to the first-order
  equation (\ref{11}), we obtain a relation connecting $b$ with $\Lambda$:
\beq
	\Lambda = 3/b^2.                               \label{Lam}
\eeq
  Noteworthy, this solution exists for $\Lambda > 0$ only.

  As a result, we obtain the following solution:
\bearr                                                     \label{ds-h}
	ds^2 = \frac{9}{\cosh^4 bu} dt^2 - b^4 du^2 -
	\frac{b^2 \cosh^2 bu}{3} d\Omega^2,
\yyy
	\phi = \pm \frac{4}{F_0}\biggl( 3u - \frac{2}{b}\tanh bu\biggr)
	+ \phi_0,
\ear
   with $\phi_0 = \const$.
  (We can remark here that there is no analytical solution for
  $\Lambda = 0$, unless the space-time signature is $(+,+,-,-)$.)

  To study the metric, it makes sense to pass on again to
  the quasiglobal coordinate $x = 3b \tanh bu$, and the metric becomes
\bearr                                                       \label{ds-q}
	ds^2 =
  	\frac{(9b^2 - x^2)^2}{9b^4} dt^2 - \frac{9b^4}{(9b^2 - x^2)^2}dx^2
\nnn \inch
		- \frac{3b^4}{9b^2 - x^2}d\Omega^2.
\ear

  It is clear from (\ref{ds-q}) that the space-time has two second-order
  (degenerate) horizons at $x = \pm 3b = \pm \sqrt{27/\Lambda}$
  with zero surface gravity (hence zero Hawking temperature), and the area
  of the horizons is infinite, that is, the horizons are of the same kind as
  have been obtained in cold black hole solutions \cite{k1, k2}. In
  particular, the tidal forces acting on extended bodies are infinite at
  horizon crossing, so only strictly point particles can cross such a
  horizon safely.

  Thus the above solution has much in common with the cold black hole
  solutions found in scalar-tensor theories \cite{k1, k2}. However, it
  cannot be called a black hole because in this space-time (in the static
  region $|x| < 3b$) there is no place for a distant observer; on the other
  hand, the existence of both horizons is connected with the cosmological
  constant $\Lambda$, hence they are cosmological in nature, similarly to
  the horizon in de Sitter space-time.

  It can be stated that these two horizons of infinite area are connected by
  a wormhole whose throat (the minimum of $r(x)$) is located on the sphere
  $x=0$. Moreover, the source of gravity, i.e., the k-essence field violates
  the NEC not only at the throat and its neighborhood (as is necessary for
  static wormholes in GR) but in the whole range of $x$, as follows from the
  inequality $d^2 r/dx^2 > 0$ (see the end of Section 2).

  Since at $x = \pm 3b$ the coefficient of $d\Omega^2$ changes its sign, the
  regions $|x| > 3b$ have the signature $(+\ -\ +\ +)$ instead of the
  original signature $(+\ -\ -\ -)$. Thus, even though the horizon is
  of even (second) order, beyond the horizon the former spatial coordinate
  $x$ becomes temporal, and so the regions $|x| > 3b$ represent anisotropic
  (Kantowski-Sachs) cosmological models like the inner region of the
  Schwarzschild space-time. There occur cosmological singularities
  at $x = \pm \infty$, and their properties also resemble the properties of
  the Schwarzschild singularity $r=0$: an extended test body is there
  squeezed to a point in the angular directions but is infinitely stretched
  in the third spatial direction corresponding to the coordinate that
  represented time in the static region $|x| < 3b$. Moreover, one can verify
  that the singularities $x = \pm \infty$ are accessed by test bodies at
  their finite proper times.

  The global structure of a space-time with the metric (\ref{ds-q}) is shown
  in Fig.\,2.
\begin{figure}[h]
\centering
\includegraphics[width=7cm]{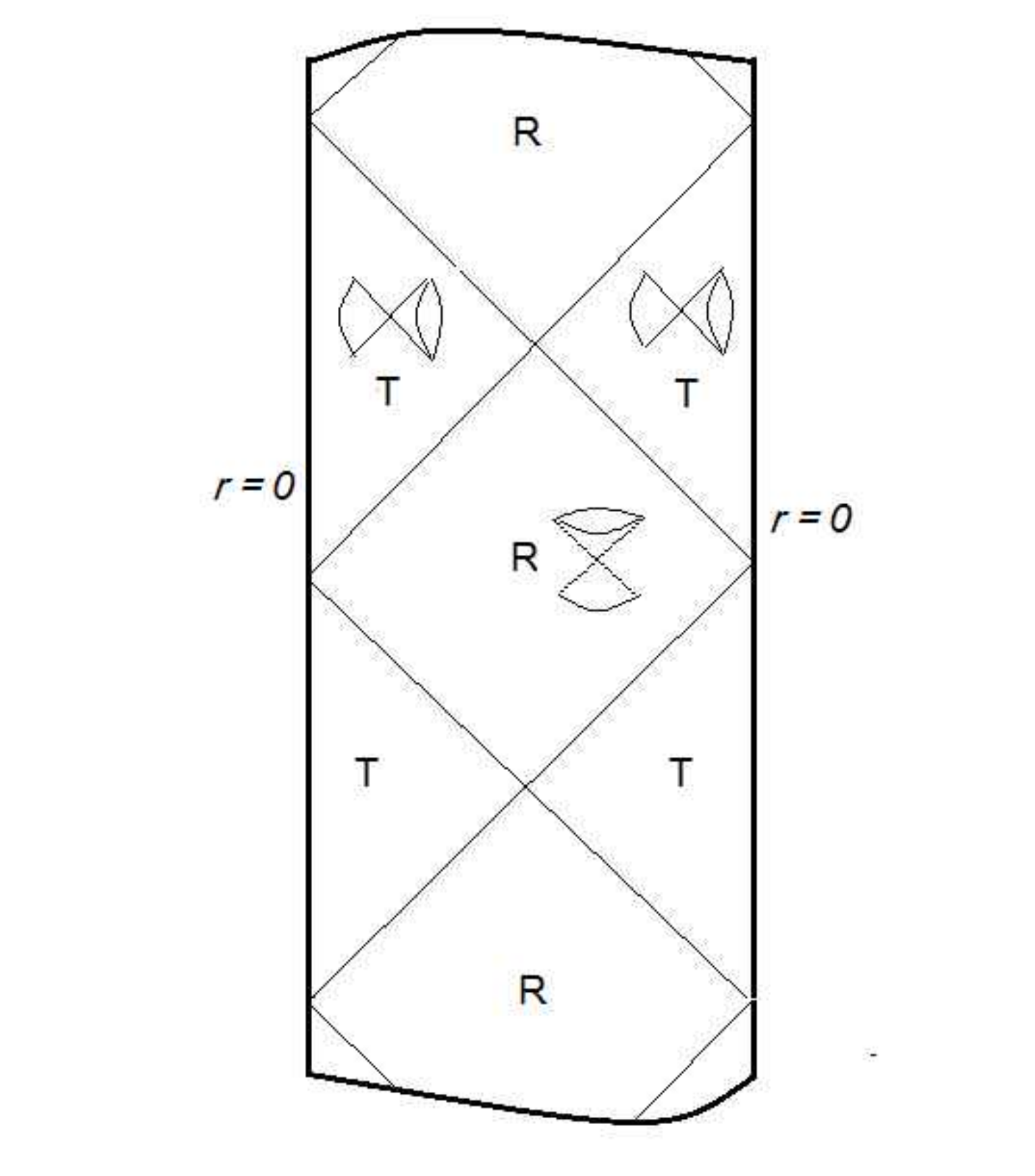}
\caption{\small
	Carter-Penrose diagram for solution 2. Light cones are shown,
	indicating the temporal direction.} \label{f2}
\end{figure}

  The scalar field $\phi$ in the whole range of $x$ is found as
\beq
      \phi = \pm \frac{4}{3F_0 b^2}
	\biggl(2x + \frac{9b}{2}\ln \Big|\frac{x-3b}{x+3b}\Big|\biggr)
	 + \phi_0
\eeq
  and is singular both at $x= \pm \infty$ and at the horizons $x=\pm 3b$,
  while $X = A\phi'^2$ is infinite at $x= \pm \infty$ and finite at the
  horizons. We thus have one more example of a horizon where the space-time
  is nonsingular but the scalar field is infinite.

\section{Conclusion}

  We have made an attempt to study \ssph\ configurations in the
  context of k-essence theories defined by a general function $F(X,\phi)$,
  where $X$ is the usual kinetic term of a scalar field. The k-essence
  theories have been employed in inflationary and dark energy models, but
  little has been done concerning local objects like stars and black
  holes. This work intended to partly fill this gap.

  We have proved a no-go theorem for the case where the k-essence theory has
  only $X$-dependence, $F(X,\phi) = f(X)$, stating that no black hole
  solution with finite horizon area is possible unless the derivative of the
  function $f(X)$ diverges at the horizon. A special solution has been
  found illustrating this no-go theorem: fixing $f(X) = F_0 X^n$ we have
  found for $n = 1/3$ a non-asymptotically flat solution, with a
  single horizon at which $f_X$ diverges while $X$ and $f$ are finite. The
  resulting configuration may be characterized as a black hole with a
  Schwarzschild-like interior immersed in an asymptotically singular
  space-time.

  Another solution has been obtained by choosing $n = 1/2$ and introducing a
  cosmological constant. This solution is also non-asymptotically flat, but
  now there are two horizons with infinite surface area. In this case,
  the function $f(X)$ is regular at the horizons but the scalar field $\phi$
  diverges there. Beyond the horizons the space-time changes its signature
  ($-2 \mapsto +2$) still remaining Lorentzian, and the solution describes
  there a Kantowski-Sachs anisotropic universe with singularities that can
  be reached by test bodies in finite proper time. Between the two horizons
  there is a static region with a wormhole geometry.

  Both solutions exemplify situations where a violent behavior of a scalar
  field at a horizon still leaves its SET finite and regular, which in turn
  leads to a regular space-time geometry. We can recall just one such
  example in the literature, a black hole with a massless conformal scalar
  field \cite{bbm} and its electrically charged counterpart \cite{kb-73}.

  It is desirable to study the stability of such solutions, in particular,
  the static region of solution 2. The stability conditions are a very
  important aspect of any black hole or wormhole type solution, which
  remains an especially interesting problem in the cases where scalar fields
  and wormhole throats are present \cite{gonz, k-z}.

\subsection*{Acknowledgments}
 
 We are grateful to Nail Khusnutdinov and Carlos Herdeiro for helpful
 discussions.
 We thank FAPES (Brazil) and CNPq (Brazil) for partial financial support.
 The work of KB was partly performed within the framework of the Center 
 FRPP supported by MEPhI Academic Excellence Project 
 (contract ¹ 02.à03.21.0005, 27.08.2013).

\end{document}